\documentclass[aps,reprint, amsmath,amssymb,aps,superscriptaddress,longbibliography,prx]{revtex4-2}

\usepackage{graphicx}
\usepackage[utf8]{inputenc}
\usepackage[T1]{fontenc}
\usepackage{xr-hyper}
\usepackage[colorlinks, allcolors=blue]{hyperref}
\usepackage{braket}
\usepackage{tikz}

\DeclareMathOperator{\Tr}{Tr}
\usepackage{physics}
\usepackage{natbib}
\setcitestyle{numbers,square}

\definecolor{mycolor}{rgb}{0.8,0.0,0.0}
\makeatletter
\newcommand*{\addFileDependency}[1]{
  \typeout{(#1)}
  \@addtofilelist{#1}
  \IfFileExists{#1}{}{\typeout{No file #1.}}
}
\makeatother
\usepackage{orcidlink}
\newcommand{\orcidalberto}{\orcidlink{0000-0002-6643-0738}}
\newcommand{\unipd}{Dipartimento di Fisica e Astronomia "G. Galilei" \& Padua Quantum Technologies Research Center, Universit{\`a} degli Studi di Padova, Italy I-35131, Padova, Italy}
\newcommand{\pdinfn}{INFN, Sezione di Padova, via Marzolo 8, I-35131, Padova, Italy}
\begin{document}

\title{Numerically efficient unitary evolution for Hamiltonians beyond nearest neighbors}

\author{Alberto Giuseppe Catalano\orcidalberto}
\email{albertogiuseppe.catalano@unipd.it}
\affiliation{Institut Ru\dj er Bo\v{s}kovi\'c, Bijeni\v{c}ka cesta 54, 10000 Zagreb, Croatia}
\affiliation{University of Strasbourg and CNRS, CESQ and ISIS (UMR 7006), aQCess, 67000 Strasbourg, France}
\affiliation{\unipd}
\affiliation{\pdinfn}
\preprint{RBI-ThPhys-2024-02}

\begin{abstract}
Matrix product states (MPSs) and matrix product operators (MPOs) are fundamental tools in the study of quantum many-body systems, particularly in the context of tensor network methods such as Time-Evolving Block Decimation (TEBD). However, constructing compact MPO representations for Hamiltonians with interactions beyond nearest-neighbors, such as those arising in atomic, molecular, and optical (AMO) systems or in systems with ring geometry, remains a challenge. 

In this paper, we propose a novel approach for the direct construction of compact MPOs tailored specifically for the exponential of spin Hamiltonians.
This approach allows for a more efficient time evolution, using TEBD, of spin systems with interactions beyond nearest-neighbors, such as long-range spin-chains, periodic systems and more complex cluster model, with interactions involving more than two spins. 
\end{abstract}
\maketitle
\section{Introduction}
Long-range spin chains, distinguished by interactions extending beyond nearest neighbors, provide a fertile ground for delving into exotic quantum phenomena \cite{Defenu2024}.
A growing interest in quantum many-body physics featuring long-range interactions is propelled by the expanding capabilities in controlling and manipulating atomic, molecular, and optical (AMO) systems. Presently, diverse platforms such as Rydberg atoms, dipolar quantum gases, polar molecules, quantum gases within optical cavities, and trapped ions exhibit inherent two-body long-range interactions characterized by algebraic decay with distance~\cite{Britton2012,Scholl2021,Monroe2021,Keesling2019,Li2023}. Unraveling the dynamic evolution of these systems presents a significant challenge, requiring theoretical frameworks that capture the interplay between quantum entanglement and long-range interactions.~\cite{Defenu2023_Rev}.

After the triumph of the density-matrix renormalization group (DMRG)~\cite{White1992,Catarina2023} in uncovering ground states of one-dimensional (1D) systems, several closely linked techniques have emerged to investigate the dynamic features of short-ranged 1D systems~\cite{Paeckel2019}.
In their recent formulation, these techniques work in the framework of matrix
product states (MPSs)~\cite{Fannes1992,Ostlund1995,Schollwock2011,Silvi2017,Montangero2018,Jaschke2019}, an efficient representation of finitely
entangled states as the product of rank-$3$ tensors, and matrix product
operators (MPOs)~\cite{Pirvu2010}, which represent quantum operators as the product of rank-$4$ tensors. If an Hamiltonian possesses a compact MPO representation
for the corresponding time evolution operator $U(t)=e^{-itH}$, meaning that the bond dimension linking the tensors in the MPO is suffciently small, then the time evolution can be efficiently simulated
by repeated application of this MPO to the MPS.
This is indeed the case in some simple systems, such as those characterized only by nearest-neighbor interactions or whose Hamiltonian can be written as the sum of commuting terms, in which it is possible to construct compact MPOs with finite error per site. This
is the basis behind the highly successful time-evolving block
decimation (TEBD)~\cite{Vidal2004, Verstraete2004,White2004} and tDMRG~\cite{Daley2004}. The main variants of TEBD use a second-order (TEBD2) or fourth-order (TEBD4) expansion of the unitary evolution operator in the time-step. Even though TEBD4 gives a smaller error per time-step, typically TEBD2 is preferred since it requires five times less MPO-MPS contractions per time-step than TEBD4. However, these methods do not
generalize well to long-range Hamiltonians, since the bond dimension of their MPOs typically scales exponentially with the range of the interaction. 

In order to overcome this issue, recently new approaches have been developed that can be applied directly to long-range Hamiltonian \cite{Zaletel2015,Feiguin2005,Ronca2017,Hauke2013,Haegeman2011,Seitz2023}. The $W^{I,II}$ methods \cite{Zaletel2015} work similarly to TEBD in the sense that they try to approximate the time evolution method for a small time step, with the advantage of producing MPOs which are usually more compact than those produced by TEBD. However, a downside of these methods is that their dynamics is not strictly unitary. Other techniques, such as the local Krylov method~\cite{Feiguin2005,Ronca2017} and the time dependent variational principle (TDVP)~\cite{Haegeman2011,Hauke2013,Secular2020} move from the standard tensor network paradigm of applying an MPO to an MPS, and try to directly approximate the time evolved state without explicitely applying the time evolution MPO to an MPS. One of the advantages of these methods is that they allow to reduce the error per time step, and in its two site variant TDVP has been shown to be the best algorithm in terms of physical accuracy and performance, the latter being comparable to that of TEDB for larger time steps~\cite{Schollwock2011}. As a drawback, common to any variational approach, in certain situations TDVP could get stuck in a local minimum, failing to converge to the exact result. 

While in many situations TDVP can outperform TEBD techniques, in this manuscript we will focus on the latter. The reason behind this choice is that TEBD is still one of the easiest methods to implement on small scale simulations and, at the same time, it ensures convergence to correct results without getting stuck in local minima, providing a solid algorithm to benchmark TDVP simulations. Within this framework, our goal is to improve the performances of TEBD in the simulation of the time evolution of systems with interactions beyond nearest-neighbors, proposing an alternative approach for the construction of MPOs for the exponentials of non-local spin operators, based on the direct exponentiation of Pauli strings. This provides a very intuitive and natural way of constructing MPOs which contain only tensors acting on a single site. The maximal bond dimension of such MPOs scales as $2^r$ with the range of the interaction $r$, as opposite to standard approaches which prescribe the application of swap gates on local two-qubits MPOs to reconstruct the desired long-range nature of the operators~\cite{Schollwock2011}, where in the typical scenario the bond dimension scales as $2^{4r-3}$. 
Therefore, while unable to cure the exponential growth of the bond dimension with the interaction range, our method still renders the TEBD technique more efficient in terms of MPO's bond dimension to simulate the time evolution of MPSs, at least for systems in which the range of the interaction is not too large. This could be the case, for example, of one dimensional Rydberg atoms systems in which, because of the $R^{-6}$ decay of the Van der Waals interactions, often only nearest and next-to-nearest neighbors interaction give a significant contribution~\cite{Schauss2018}. 
Our approach is also relevant to the study of short-range spin chains with periodic boundary conditions, in which a single operator with non-local structure emerges at the boundaries of the system. Indeed, while in Hamiltonians with non-local interactions and open boundary conditions one could try to reduce the overhead introduced by the SWAP operators with special rearrangements~\cite{Schollwock2011}, in the presence of a single non-local interaction connecting the first and last spin in the chain this is not possible. In this case, our method produces an MPO with constant bond dimension $w=4$, which is eight times smaller than the bond dimension $w=32$ achieved with the SWAP gates. This would significantly improve the performance of the simulations of the dynamics of those systems which are very sensitive to the presence of periodic boundary conditions. This is the case, for example, of ring-shaped networks of Rydberg atoms, which exhibit interesting transport properties~\cite{Perciavalle2024,Perciavalle2023_2,Polo2020,Catalano2023-LR}, or of topologically frustrated spin chains, in which the combination of short-range antiferromagnetic interactions, odd number of spins and periodic boundary conditions has been proven to produce very interesting consequences, such as the modification of order paramaters~\cite{Maric2020_destroy, Maric2020_induced}, the closure of the energy gap in typically gapped phase~\cite{Catalano2022,Sacco2023}, the presence of unexpected long-range correlations~\cite{Torre2023} and a complex dynamics, as witnessed by the Loschmidt echo~\cite{Torre2022}. Moreover, our technique is easily generalized also to Hamiltonians with more complex cluster interactions~\cite{Smacchia2011,Giampaolo2014,Zonzo2018,Perk2017,Ding2019}, i.e. interactions involving more than two spins. 

The manuscript is organized as follows. After discussing the standard TEBD approach in Sec.~\ref{sec:tebd}, we will introduce our technique for the efficient MPO representation of the exponentials of spin Hamiltonians in Sec.~\ref{sec:new_method}. In Sec.~\ref{sec:application}, to benchmark our method, we study the dynamical properties of some nonintegrable models, namely the axial next nearest neighbors Ising (ANNNI) chain and the two-dimensional quantum Ising model on a cylinder. Finally, we discuss our results in Sec.~\ref{sec:conclusions}.

\section{Standard TEBD and its problems for long-range systems}
\label{sec:tebd}
In order to present the general ideas behind TEBD and later, in Sec.~\ref{sec:new_method}, our novel approach, it is sufficient to start by considering the following family of Hamiltonians describing open spin chains with long-range interactions 
\begin{equation}
    \label{eq:model}
    H_r(J,h)=J\sum_{l=1}^{N-r}\sigma_l^x\sigma_{l+r}^x + h\sum_{l=1}^N\sigma_l^z = J X_{r,N} + hZ_{N}, 
\end{equation}
where $r<N/2$ is the range of the interaction along the $\vec{x}$ direction, $J$ determines its nature and strength, $h$ is a transverse magnetic field along the $\vec{z}$ direction and $\sigma^\alpha$ for $\alpha=0,x,y,z$ are the Pauli matrices.

At its heart, TEBD relies on a Trotter-Suzuki decomposition~\cite{Suzuki1976} to approximate the time-evolution operator $U(\delta)$. Using the Hamiltonian \eqref{eq:model} as an example, this decomposition gives
\begin{align}
\label{eq:Trotter}
    U(\delta)&=e^{-i\delta H_r}\approx e^{-i\frac{h\delta}{2}Z_{N}}e^{-iJ\delta X_{r,N}}e^{-i\frac{h\delta}{2}Z_{N}} + o(\delta^3)\nonumber\\&=U^{TEBD2}(\delta),
\end{align}
where every exponential appearing in \eqref{eq:Trotter} is made up by commuting terms, e.g. $e^{-i J\delta X_{r,N}}=\prod_{l=1}^{N-r}e^{-iJ\delta \sigma_l^x\sigma_{l+r}^x}$.
If we imagine to evolve the system over a time interval $T$ which we divide in $T/\delta$ steps, replacing the exact time evolution operator $U(\delta)$ with $U^{TEDB2}(\delta)$ yields an error of order $\delta^2$ after every time interval of length $T$. 

After the decomposition is chosen, the application of tensor network techniques to time evolve an MPS requires the construction of MPOs for the single and two-qubit gates appearing in \eqref{eq:Trotter}. Since the construction of the first ones is trivial and does not produce any overhead in terms of bond dimension, here we will focus only on the construction of the MPO representation of two-qubit gates. 

Let $E_{l,l+r}=e^{-iJ\delta \sigma_l^x\sigma_{l+r}^x}$. In presence of short-range interactions, i.e. for $r=1$, one can start from an MPO of length $N-1$ containing a tensor with four physical legs at site $l$ and bond dimension $w=1$ at every link (as indicated above the legs connecting neighbors tensors):
\begin{equation}
\label{eq:E_1}
\resizebox{0.9\linewidth}{!}{
    \begin{tikzpicture}[every node/.style={font=\huge}]
    \node[minimum size=1cm, draw, shape=circle] (v0) at (0,0) {$\mathbb{I}$};
    \node[minimum size=1cm,draw, shape=circle] (v1) at (1.5,0) {$\mathbb{I}$};
    \node[minimum size=1cm,draw, shape=circle] (v2) at (4.5,0) {$\mathbb{I}$};
    \node[minimum size=1cm,draw, shape=circle] (v3) at (6.,0) {$Q$};
    \node[minimum size=1cm,draw, shape=circle] (v4) at (7.5,0) {$\mathbb{I}$};
    
    \node[minimum size=1cm,draw, shape=circle] (v6) at (10.5,0) {$\mathbb{I}$};
    \draw [thick] (v0) -- (v1)
    (v3) -- (v4)
    (v2) -- (v3)
    (v0) -- (0,-1) 
    (v0) -- (0,1) 
    (v0) -- (-1,0)
    (v1) -- (1.5,1)
    (v6) -- (9.5,0)
    (v1) -- (2.5,0)
    (v2) -- (4.5,1)
    (v2) -- (4.5,-1)
    (6.25,0.53301) -- (6.25,1)
    (5.75,0.53301) -- (5.75,1)
    (6.25,-0.53301) -- (6.25,-1)
    (5.75,-0.53301) -- (5.75,-1)
    (v4) -- (7.5,1)
    (v4) -- (7.5,-1)
    (v4) -- (8.5,0);

    \draw[dashed] (2.55,0) -- (3.45,0)
    (8.55,0) -- (9.45,0);
    \draw[thick] (3.5,0) -- (v2)
    (v1) -- (1.5,-1)
    (v6) -- (10.5,1)
    (v6) -- (10.5,-1)
    (v6) -- (11.5,0);
    \node[] at (.75,.35) {$1$};
    \node[] at (2.25,.35) {$1$};
    \node[] at (3.75,.35) {$1$};
    \node[] at (5.25,.35) {$1$};
    \node[] at (6.75,.35) {$1$};
    \node[] at (8.25,.35) {$1$};
    \node[] at (9.75,.35) {$1$};
    \node[] at (11.25,.35) {$1$};
    \node[] at (-2.5,0) {$E_{l,l+1}=$};
    \end{tikzpicture}}
\end{equation}
where $Q=e^{-i\delta J \sigma^x\otimes\sigma^x}$ is the local two-qubit gate of interest and $\mathbb{I}$ is the $2\times2$ identity matrix. At this point, in order to complete the construction of our MPO we can split the tensor $Q$ in two tensors, e.g. by SVD, and obtain the desired $N$-length MPO

\begin{equation}
\resizebox{0.9\linewidth}{!}{
\begin{tikzpicture}[every node/.style={font=\huge}]
    \tikzset{
        fixedcircle/.style={
            draw, circle, minimum size=1.2cm, inner sep=0pt, align=center
        }
    }

    \node[fixedcircle] (v0) at (0,0) {\makebox[1cm][c]{$\mathbb{I}$}};
    \node[fixedcircle] (v1) at (1.7,0) {\makebox[1cm][c]{$\mathbb{I}$}};
    \node[fixedcircle] (v2) at (4.5,0) {\makebox[1cm][c]{$\mathbb{I}$}};
    \node[fixedcircle] (v3) at (6.2,0) {\makebox[1cm][c]{\LARGE $q_l$}};
    \node[fixedcircle] (v4) at (7.9,0) {\makebox[1cm][c]{\LARGE $q_{l+1}$}};
    \node[fixedcircle] (v5) at (9.6,0) {\makebox[1cm][c]{$\mathbb{I}$}};
    \node[fixedcircle] (v6) at (12.4,0) {\makebox[1cm][c]{$\mathbb{I}$}};

    \draw [thick] 
        (v0) -- (v1)
        (v1) -- (2.7,0)
        (v2) -- (v3)
        (v3) -- (v4)
        (v4) -- (v5)
        (v5) -- (10.5,0)
        (v0) -- (0,-1)
        (v0) -- (0,1)
        (v0) -- (-1,0)
        (v1) -- (1.7,1)
        (v1) -- (1.7,-1)
        (v2) -- (4.5,1)
        (v2) -- (4.5,-1)
        (v3) -- (6.2,1)
        (v3) -- (6.2,-1)
        (v4) -- (7.9,1)
        (v4) -- (7.9,-1)
        (v5) -- (9.6,1)
        (v5) -- (9.6,-1)
        (v6) -- (12.4,1)
        (v6) -- (12.4,-1)
        (v6) -- (13.4,0)
        (v6) -- (11.4,0)
        (v2) -- (3.5,0);

    \draw[dashed] (2.55,0) -- (3.45,0)
                  (10.5,0) -- (11.45,0);

    \node at (.85,.35) {$1$};
    \node at (2.5,.35) {$1$};
    \node at (3.75,.35) {$1$};
    \node at (5.35,.35) {$1$};
    \node at (7.05,.35) {{\Large $m$}};
    \node at (8.75,.35) {$1$};
    \node at (10.35,.35) {$1$};
    \node at (11.6,.35) {$1$};

    \node at (-2.5,0) {$E_{l,l+1}=$};
\end{tikzpicture}
}
\end{equation}

which will have bond dimension $w=1$ at every site but between site $l$ and $l+1$ where a bond has been created splitting the two-qubit gate and $w=m$. 
Therefore, it is easy to understand that the full exponential $e^{-iJ\delta X_{1,N}}=\prod_{l=1}^{N-1}E_{l,l+1}$ will be represented by an MPO with constant bond dimension $w=m$ at every site.  

For long-range systems, i.e. for $r>1$, the standard prescription to build an MPO for $E_{l,l+r}$ is to start from the local two-qubit operator $E_{l,l+1}$ and apply swap operators to obtain the desired non-local structure of the operator~\cite{Schollwock2011}. Therefore, if we denote by $S_{l,l+1}$ the swap gate between site $l$ and $l+1$, we have that
\begin{equation}
E_{l,l+r}=\mathcal{S}_{l,r}E_{l,l+1}\mathcal{S}_{l,r}^\dagger,
\end{equation}
where $\mathcal{S}_{l,r}=\prod_{k=l+1}^{l+r-1}S_{k,k+1}$. To understand the resulting MPO structure for $E_{l,l+r}$ we should first take into account the structure of the MPO representing the swap operators. This is given by
\begin{equation}
\label{eq:swap_MPO}
\resizebox{0.9\linewidth}{!}{
\begin{tikzpicture}[every node/.style={font=\huge}]
    \tikzset{
        fixedcircle/.style={
            draw, circle, minimum size=1.2cm, inner sep=0pt, align=center
        }
    }

    \node[fixedcircle] (v0) at (0,0) {\makebox[1cm][c]{$\mathbb{I}$}};
    \node[fixedcircle] (v1) at (1.7,0) {\makebox[1cm][c]{$\mathbb{I}$}};
    \node[fixedcircle] (v2) at (4.5,0) {\makebox[1cm][c]{$\mathbb{I}$}};
    \node[fixedcircle] (v3) at (6.2,0) {\makebox[1cm][c]{\LARGE $s_k$}};
    \node[fixedcircle] (v4) at (7.9,0) {\makebox[1cm][c]{\LARGE $s_{k+1}$}};
    \node[fixedcircle] (v5) at (9.6,0) {\makebox[1cm][c]{$\mathbb{I}$}};
    \node[fixedcircle] (v6) at (12.4,0) {\makebox[1cm][c]{$\mathbb{I}$}};

    \draw [thick] 
        (v0) -- (v1)
        (v1) -- (2.7,0)
        (v2) -- (v3)
        (v3) -- (v4)
        (v4) -- (v5)
        (v5) -- (10.5,0)
        (v0) -- (0,-1)
        (v0) -- (0,1)
        (v0) -- (-1,0)
        (v1) -- (1.7,1)
        (v1) -- (1.7,-1)
        (v2) -- (4.5,1)
        (v2) -- (4.5,-1)
        (v3) -- (6.2,1)
        (v3) -- (6.2,-1)
        (v4) -- (7.9,1)
        (v4) -- (7.9,-1)
        (v5) -- (9.6,1)
        (v5) -- (9.6,-1)
        (v6) -- (12.4,1)
        (v6) -- (12.4,-1)
        (v6) -- (13.4,0)
        (v6) -- (11.4,0)
        (v2) -- (3.5,0);

    \draw[dashed] (2.55,0) -- (3.45,0)
                  (10.5,0) -- (11.45,0);

    \node at (.85,.35) {$1$};
    \node at (2.5,.35) {$1$};
    \node at (3.75,.35) {$1$};
    \node at (5.35,.35) {$1$};
    \node at (7.05,.35) {{$4$}};
    \node at (8.75,.35) {$1$};
    \node at (10.35,.35) {$1$};
    \node at (11.6,.35) {$1$};

    \node at (-2.5,0) {$S_{k,k+1}=$};
\end{tikzpicture}
}
\end{equation}
where the tensors at site $k$ and $k+1$ are a row and a column vector, respectively
\begin{equation}
\label{eq:mpo_blocks}
    s_k=\begin{bmatrix} 
\mathbb{I}, & \sigma^{x}, & \sigma^{y}, & \sigma^z  \\ 
\end{bmatrix}, \;\; s_{k+1}=\begin{bmatrix} 
\mathbb{I}/2 \\ 
\sigma^x/2  \\ \sigma^y/2 \\ \sigma^z/2 
\end{bmatrix}.
\end{equation}
Accordingly, the swap strings $\mathcal{S}_{l,r}$ will be represented by an MPO which has bond dimension $w=4$ at all sites between $l+1$ and $l+r$. Therefore, the resulting MPO for $E_{l,l+r}$ built using the swap operators will have bond dimension $w=q$ between site $l$ and $l+1$ and then constant bond dimension $w=16$ at every site between sites $l+1$ and $l+r$
\begin{equation}
\label{eq:E_r}
\resizebox{0.9\linewidth}{!}{
\begin{tikzpicture}[every node/.style={font=\huge}]
    \tikzset{
        fixedcircle/.style={
            draw, circle, minimum size=1cm, inner sep=0pt, align=center
        }
    }

    \node[fixedcircle] (v0) at (0,0) {\makebox[1cm][c]{$\mathbb{I}$}};
    \node[fixedcircle] (v2) at (3,0) {\makebox[1cm][c]{$e_l$}};
    \node[fixedcircle] (v3) at (4.5,0) {\makebox[1cm][c]{\LARGE $e_{l+1}$}};
    \node[fixedcircle] (v4) at (6.05,0) {\makebox[1cm][c]{\LARGE $e_{l+2}$}};
    \node[fixedcircle] (v5) at (9.,0) {\makebox[1cm][c]{\LARGE $e_{l+r}$}};
    \node[fixedcircle] (v6) at (12,0) {\makebox[1cm][c]{$\mathbb{I}$}};

    \draw [thick]
        (v0) -- (1,0)
        (v2) -- (v3)
        (v3) -- (v4)
        (v4) -- (7,0)
        (v5) -- (9,1)
        (v5) -- (9,-1)
        (v5) -- (10,0)
        (v5) -- (8,0)
        (v4) -- (6.05,1)
        (v4) -- (6.05,-1)
        (v3) -- (4.5,1)
        (v3) -- (4.5,-1)
        (v2) -- (3,1)
        (v2) -- (3,-1)
        (v2) -- (2,0)
        (v0) -- (0,-1)
        (v0) -- (0,1)
        (v0) -- (-1,0)
        (v6) -- (11,0)
        (v6) -- (12,1)
        (v6) -- (12,-1)
        (v6) -- (13,0);

    \draw[dashed] 
        (1.05,0) -- (1.95,0)
        (7.05,0) -- (7.95,0)
        (10,0) -- (10.95,0);

    \node at (.75,.35) {$1$};
    \node at (2.25,.35) {$1$};
    \node at (3.75,.35) {$q$};
    \node at (5.35,.35) {\LARGE $2^4$};
    \node at (6.9,.35) {\LARGE $2^4$};
    \node at (8.3,.35) {\LARGE $2^4$};
    \node at (9.75,.35) {$1$};
    \node at (11.25,.35) {$1$};

    \node at (-2.5,0) {$E_{l,l+r}=$};
\end{tikzpicture}}
\end{equation}

Therefore, we can now easily understand that the MPO for the full exponential $e^{-iJ\delta X_{r,N}}=\prod_{l=1}^{N-r}E_{l,l+r}$ will have a maximum bond dimension of $w=m\cdot2^{4r-4}$, producing a minimum scaling which goes like $2^{4r-3}$ when $m=2$. This obviously scales quite badly with the range of the interaction, making the TEBD method quite inefficient for long-range systems. 

Similarly, it is easy to understand that in the case of an Hamiltonian with nearest-neighbors interactions ($r=1$) and periodic boundary conditions the total bond dimension would be $w=m\cdot 16$, since the number of swap operators needed to build $E_{1,N}$ is of the order of the system's size. 

In the next section we will propose a different approach to the construction of the MPOs for $E_{l,l+r}$, which does not require the application of any SWAP operator.

\section{Compact MPOs for the exponential of Pauli strings}
\label{sec:new_method}
Let $\boldsymbol{\sigma}=\bigotimes_{i=1}^N \sigma^{\alpha_i}$ be a $N$-qubit Pauli string, where $\sigma^\alpha$ for $\alpha=0,x,y,z$ are Pauli matrices. We are interested in the evaluation of the exponential
$e^{-i \delta \boldsymbol{\sigma}}$. This becomes quite straightforward if one considers that $\boldsymbol{\sigma}^2=\mathbb{I}^{\otimes N}$. Indeed, we have that
\begin{align}
    e^{-i \delta \boldsymbol{\sigma}}&= \sum_{n=0}^{\infty}\frac{(-i\delta)^n}{n!}\boldsymbol{\sigma}^n\nonumber\\&=\sum_{n=0}^{\infty}\frac{(-1)^n\delta^{2n}}{(2n)!}\mathbb{I}^{\otimes N}-i\sum_{n=0}^{\infty}\frac{(-1)^n\delta^{2n+1}}{(2n+1)!}\boldsymbol{\sigma},
\end{align}
which yields to
\begin{equation}
    e^{-i \delta \boldsymbol{\sigma}}=\cos(\delta)~\mathbb{I}^{\otimes_N}-i\sin(\delta)~\boldsymbol{\sigma}.
    \label{eq:exponential}
\end{equation}
The expression in \eqref{eq:exponential} can be easily expressed in tensor networks language as an MPO with constant bond-dimension $w=2$, where the single qubit operators are given by the rank-$4$ tensor
\begin{equation}
    O_l=\begin{bmatrix} 
\mathbb{I} & 0 \\ 
0 & \sigma^{\alpha_l}   \\ 
\end{bmatrix}, \;\;\; l=1,\dots,N-1,
\end{equation}
\begin{equation}
    O_N=\begin{bmatrix} 
\cos(\delta)~\mathbb{I} & 0 \\ 
0 & -i\sin(\delta)~\sigma^{\alpha_N}   \\ 
\end{bmatrix}.
\end{equation}

The expression of these MPOs can become even more compact when we are treating two-qubit gates. Using the notation introduced in Sec.~\ref{sec:tebd}, for local two-qubit gates we indeed have that \eqref{eq:exponential} can be expressed as 
\begin{equation}
\label{eq:E_1_pauli}
\resizebox{0.9\linewidth}{!}{
\begin{tikzpicture}[every node/.style={font=\huge}]
    \tikzset{
        fixedcircle/.style={
            draw, circle, minimum size=1.2cm, inner sep=0pt, align=center
        }
    }

    \node[fixedcircle] (v0) at (0,0) {\makebox[1cm][c]{$\mathbb{I}$}};
    \node[fixedcircle] (v1) at (1.7,0) {\makebox[1cm][c]{$\mathbb{I}$}};
    \node[fixedcircle] (v2) at (4.5,0) {\makebox[1cm][c]{$\mathbb{I}$}};
    \node[fixedcircle] (v3) at (6.2,0) {\makebox[1cm][c]{\LARGE $a_l$}};
    \node[fixedcircle] (v4) at (7.9,0) {\makebox[1cm][c]{\LARGE $a_{l+1}$}};
    \node[fixedcircle] (v5) at (9.6,0) {\makebox[1cm][c]{$\mathbb{I}$}};
    \node[fixedcircle] (v6) at (12.4,0) {\makebox[1cm][c]{$\mathbb{I}$}};

    \draw [thick] 
        (v0) -- (v1)
        (v1) -- (2.7,0)
        (v2) -- (v3)
        (v3) -- (v4)
        (v4) -- (v5)
        (v5) -- (10.5,0)
        (v0) -- (0,-1)
        (v0) -- (0,1)
        (v0) -- (-1,0)
        (v1) -- (1.7,1)
        (v1) -- (1.7,-1)
        (v2) -- (4.5,1)
        (v2) -- (4.5,-1)
        (v3) -- (6.2,1)
        (v3) -- (6.2,-1)
        (v4) -- (7.9,1)
        (v4) -- (7.9,-1)
        (v5) -- (9.6,1)
        (v5) -- (9.6,-1)
        (v6) -- (12.4,1)
        (v6) -- (12.4,-1)
        (v6) -- (13.4,0)
        (v6) -- (11.4,0)
        (v2) -- (3.5,0);

    \draw[dashed] (2.55,0) -- (3.45,0)
                  (10.5,0) -- (11.45,0);

    \node at (.85,.35) {$1$};
    \node at (2.5,.35) {$1$};
    \node at (3.75,.35) {$1$};
    \node at (5.35,.35) {$1$};
    \node at (7.05,.35) {$2$};
    \node at (8.75,.35) {$1$};
    \node at (10.35,.35) {$1$};
    \node at (11.6,.35) {$1$};

    \node at (-2.5,0) {$E_{l,l+1}=$};
\end{tikzpicture}
}
\end{equation}
where
\begin{equation}
\label{eq:mpo_blocks}
     a_l=\begin{bmatrix} 
\mathbb{I}, & \sigma^{x}   \\ 
\end{bmatrix}, \;\; a_{l+1}=\begin{bmatrix} 
\cos(J \delta)~\mathbb{I} \\ 
-i\sin(J \delta )~\sigma^{x}   \\ 
\end{bmatrix},
\end{equation}
which has bond dimension $w=1$ at every link but the one connecting the $l$-th and $(l+1)$-th qubit, where $w=2$. For non-local two qubit gates we instead have
\begin{equation}
\label{eq:E_r_pauli}
\resizebox{0.9\linewidth}{!}{
\begin{tikzpicture}[every node/.style={font=\huge}]
    \tikzset{
        fixedcircle/.style={
            draw, circle, minimum size=1cm, inner sep=0pt, align=center
        }
    }

    \node[fixedcircle] (v0) at (0,0) {\makebox[1cm][c]{$\mathbb{I}$}};
    \node[fixedcircle] (v2) at (3,0) {\makebox[1cm][c]{\LARGE $a_l$}};
    \node[fixedcircle] (v3) at (4.5,0) {\makebox[1cm][c]{$I_2$}};
    \node[fixedcircle] (v4) at (6,0) {\makebox[1cm][c]{$I_2$}};
    \node[fixedcircle] (v5) at (9,0) {\makebox[1cm][c]{\LARGE $a_{l+r}$}};
    \node[fixedcircle] (v6) at (12,0) {\makebox[1cm][c]{$\mathbb{I}$}};

    \draw [thick]
        (v0) -- (1,0)
        (v2) -- (v3)
        (v3) -- (v4)
        (v4) -- (7,0)
        (v5) -- (9,1)
        (v5) -- (9,-1)
        (v5) -- (10,0)
        (v5) -- (8,0)
        (v4) -- (6,1)
        (v4) -- (6,-1)
        (v3) -- (4.5,1)
        (v3) -- (4.5,-1)
        (v2) -- (3,1)
        (v2) -- (3,-1)
        (v2) -- (2,0)
        (v0) -- (0,-1)
        (v0) -- (0,1)
        (v0) -- (-1,0)
        (v6) -- (11,0)
        (v6) -- (12,1)
        (v6) -- (12,-1)
        (v6) -- (13,0)
        (3.5,0) -- (v2);

    \draw[dashed] 
        (1.05,0) -- (1.95,0)
        (7.05,0) -- (7.95,0)
        (10,0) -- (10.95,0);

    \node at (.75,.35) {$1$};
    \node at (2.25,.35) {$1$};
    \node at (3.75,.35) {$2$};
    \node at (5.25,.35) {$2$};
    \node at (6.75,.35) {$2$};
    \node at (8.25,.35) {$2$};
    \node at (9.75,.35) {$1$};
    \node at (11.25,.35) {$1$};

    \node at (-2.5,0) {$E_{l,l+r}=$};
\end{tikzpicture}}
\end{equation}

where
\begin{equation}
    I_2=\begin{bmatrix}
    \mathbb{I} & 0 \\
    0 & \mathbb{I}\\
    \end{bmatrix}.
\end{equation}
It is immediate to see that the MPO \eqref{eq:E_r_pauli} constructed using the direct exponentiation of Pauli strings is much more compact than \eqref{eq:E_r}. Building the full exponential matrix $\prod_{l=1}^{N-r} E_{l,l+r}$ that enters the time evolution operator, using \eqref{eq:E_r_pauli} we will thus end up with a final MPO whose maximum bond dimension is $w=2^r$. Unfortunately, this is still scaling exponentially with the range of the interaction, but it still provides a much better scaling than using swap gates. 

For systems with $r=1$ and periodic boundary conditions, we have that $E_{1,N}$ has an MPO representation with constant bond dimension $w=2$, resulting in a total bond dimension of $w=4$ when multiplied with the exponentials of the local two-qubit interactions. This is at least $8$ times smaller than the one obtained using SWAP gates.

Moreover, we would like to highlight that, thanks to \eqref{eq:exponential}, also the construction of MPOs for the exponential of Hamiltonians with more complicated cluster interactions could become quite compact and straightforward. 

\section{Some applications}
\label{sec:application}

In this section, we will use TEBD to simulate the time evolution of some non-integrable spin Hamiltonians after a global quantum quench in the external magnetic field. 
To show the versatility of our method in reproducing different observables, we will evaluate quantities such as the Loschmidt echo
\begin{equation}
    \mathcal{L}(t)=|\bra{\psi_0}\ket{\psi(t)}|^2,
\end{equation}
and the bipartite entanglement entropy 
\begin{equation}
 S_A(t) = - \Tr[\rho_A(t) \log_2 \rho_A(t)],
\end{equation}
where $A$ and $B$ are subpartitions of the total system, and $\rho_A(t)=\Tr_B(\ketbra{\psi(t)}{\psi(t)})$. 
The choice of these quantities is dictated by their relevance to the study of the dynamical properties of many-body spin chains. For example, the Loschmidt echo is known to detect dynamical quantum phase transitions (DQPTs)~\cite{Heyl2013,Andraschko2014,Heyl2018}, highlighted by non-analytical behaviors in the time evolution of its rate function 
\begin{equation}
    \label{eq:rate}
    r(t)=-\frac{1}{N}\log \mathcal{L}(t).
\end{equation}
Instead, entanglement entropy captures how quantum correlations spread and evolve through the system over time. Its growth thus reveals how information propagates after a quantum quench, how thermalization or localization emerges, and how constraints such as integrability or conservation laws shape the dynamics. This makes it an essential tool for probing fundamental aspects of non-equilibrium quantum many-body physics.

\begin{figure*}[th!]
    \centering
    \includegraphics[width=\textwidth]{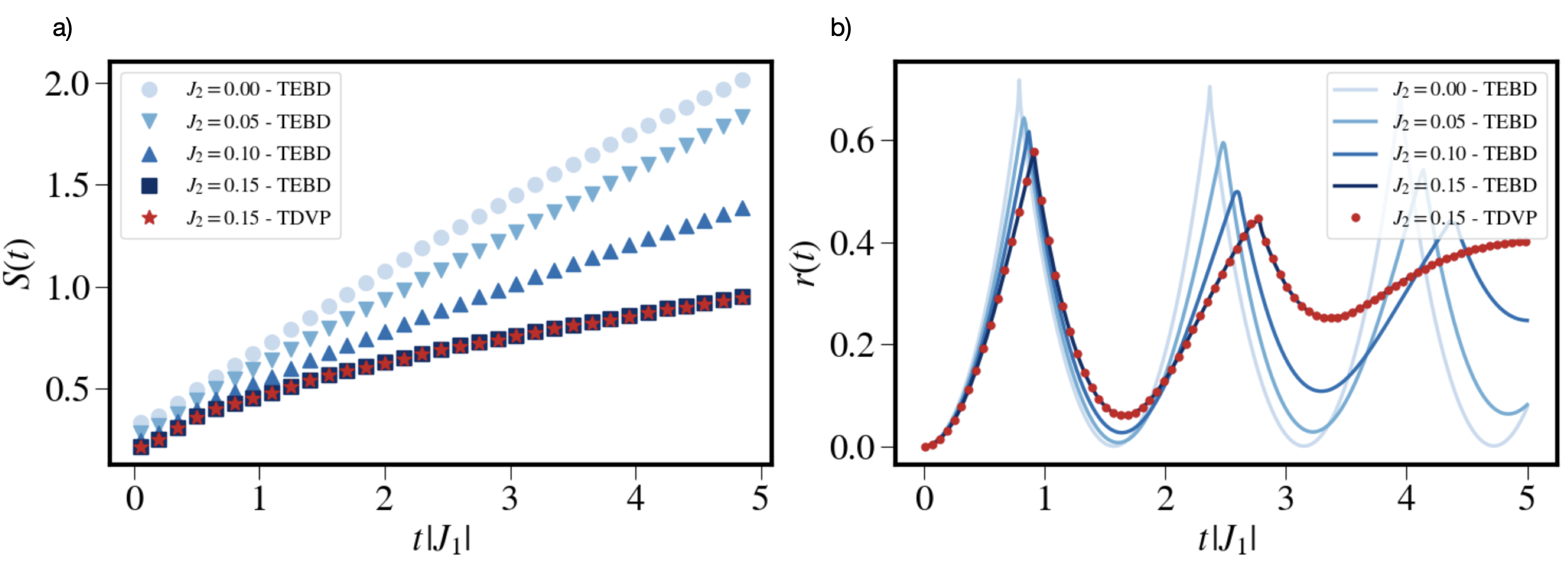}
    \caption{a) Time evolution of the half chain bipartite entanglement after global quantum quenches of the external magnetic field from the disordered phase ($h_0=1.25$) into the ordered one ($h_1=0.90$). Blue markers represent simulations carried out with our TEBD approach, while the red stars correspond to single site TDVP simulations. The data are collected for $J_1=-1$, $N=60$, $\delta=0.05$ and different values of $J_2\in[0,0.15]$. Increasing $J_2$ reduces entanglement growth, signaling that frustrated competing interactions slow the spread of correlations in the system. b) Evolution of the Loschmidt echo rate function. The data are collected for $J_1=-1$, $N=100$, $\delta=0.01$ and $J_2\in[0,0.15]$ after a global quantum quench from $h_0=2.00$ to $h_1=0.01$. The fact that $J_2$ breaks integrability is reflected in the loss of exact periodicity of $r(t)$. Moreover, the number of peaks and their sharpness is reduced with increasing frustration strength.}
    \label{fig:annni_obs}
\end{figure*}
\begin{figure*}[th!]
    \centering
    \includegraphics[width=\textwidth]{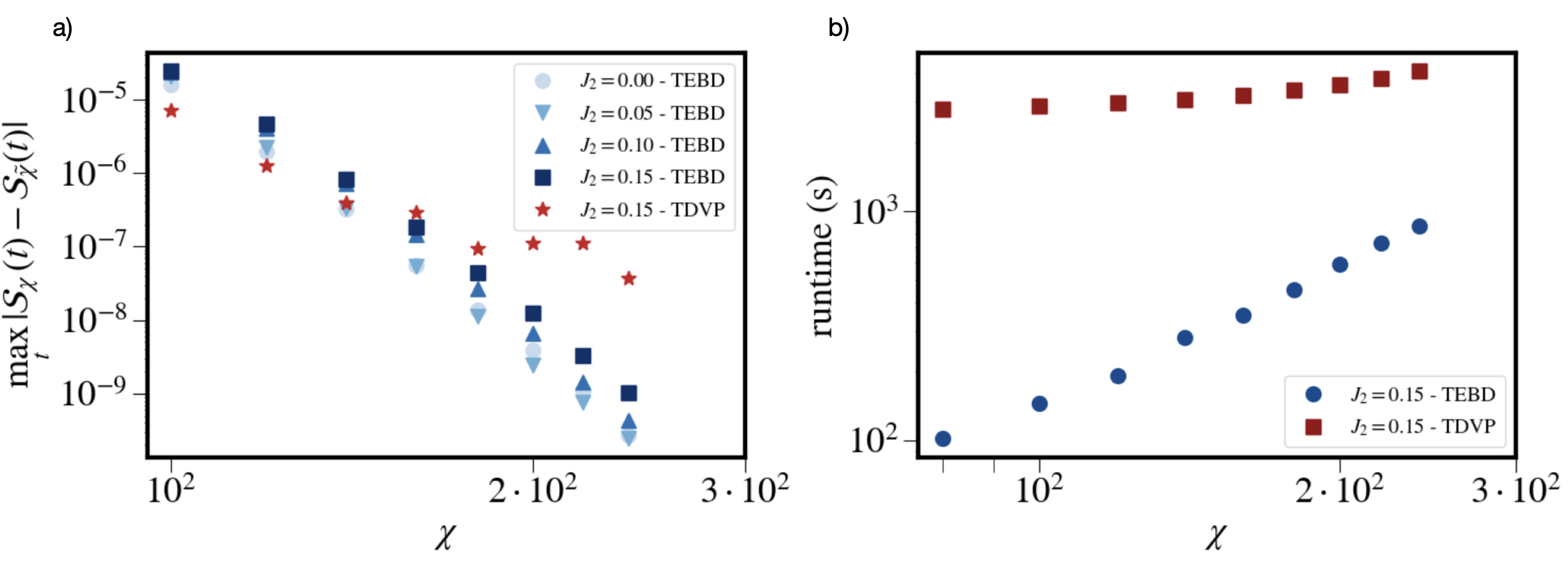}
    \caption{a) Scaling of the truncation errors with the bond dimension $\chi$ for the entanglement entropy shown in Fig.~\ref{fig:annni_obs}. b) Scaling of the numerical simulation time with $\chi$. Blue markers represent simulations carried out with our TEBD approach, while the red ones correspond to single site TDVP simulations. We can observe that, for the case studied, the performances of our approach are better than those of single site TDVP simulations.}
    \label{fig:annni_err}
\end{figure*}
\subsection{The ANNNI chain}
The first model that we examine to benchmark our approach is the quantum Ising chain with next-nearest-neighbor interactions, aka the one-dimensional ANNNI model. Its Hamiltonian reads~\cite{Selke1988}
\begin{align}
\label{eq:annni}
    H_{ANNNI}&=\sum_{\ell=1}^{N}(J_1 \sigma_{\ell}^x\sigma_{\ell+1}^x+J_2 \sigma_{\ell}^x\sigma_{\ell+2}^x)+h\sum_{\ell=1}^N \sigma_\ell^z\nonumber\\&=\sum_{r=1}^2J_rX_r + hZ_N,
\end{align}
where $J_1$ and $J_2$ are, respectively, the nearest and next-to-nearest neighbors interactions, $h$ is a transverse magnetic field and periodic boundary conditions are taken into account, i.e., $\sigma^{\alpha}_{\ell+N}=\sigma^{\alpha}_{\ell}$ for $\alpha=x,y,z$.
Such a system is generally considered as the prototypical example of frustrated quantum system. Indeed, when the interaction $J_2$ is antiferromagnetic (i.e., $J_2>0$), the competition between the different magnetic orders promoted by $J_1$ and $J_2$ produces a rich phase phase diagram, characterized by four different phases~\cite{Beccaria2006,Beccaria2007,Canabarro2019,Monaco2023}: an ordered phase, in which the ferromagnetic or antiferromagnetic order is determined by the sign of $J_1$; a paramagnetic disordered phase, for large $h$; an antiphase, dominated by the interaction $J_2>|J_1|/2$, and a floating phase separating the disordered and antiphase. In this work, we will set $J_1=-1$, and study the time evolution of $r(t)$ and $S(t)$ after global quantum quenches of the external magnetic field from the disordered phase to the ordered one, for different values of the next-nearest neighbors interaction $J_2$. 
The trotterized time-evolution operator for the Hamiltonian in Eq.~\eqref{eq:annni} can be written as
\begin{equation}
    U^{TEBD2}=e^{-i\frac{h\delta}{2}Z_N}e^{-iJ_1\delta X_{1,N}}e^{-iJ_2\delta X_{2,N}}e^{-i\frac{h\delta}{2}Z_N}.
\end{equation}
Here, the MPO representation with the highest bond dimension is the one corresponding to the operator $e^{-iJ_2\delta X_{2,N}}$. Using our technique and taking into account the periodic boundary conditions, this is equal to $w=2^4$, which is $2^5$ times smaller than the one obtained with a naive application of swap gates.

The results obtained from the simulation of the time evolution for $r(t)$ and $S(t)$ are reported in Fig.~\ref{fig:annni_obs}. For $J_2=0$, i.e., in the quantum Ising regime, we reproduced a very well known result, showing that the half chain bipartite entanglement entropy displays a linear growth in time. In particular, the slope of $S(t)$ has been connected to the propagation velocity of quasiparticle excitations along the chain after a global quantum quench. When we introduce weak frustration, i.e., turn on a small $J_2>0$, we see that this slope is reduced, signaling that the presence of more complex frustrated interactions slows the spread of correlations throughout the system. 
Also the behavior of the rate function is influenced by the introduction of frustration in the system. While for $J_2=0$ DQPTs are marked by periodic sharp peaks of $r(t)$, when $J_2>0$ the periodicity is loss, signaling the breaking of integrability, and the peaks become less and less sharp. These results are in agreement with those reported in~\cite{Cheraghi2020,Karrasch2013}, and with the results obtained by simulating the time evolution with a single site TDVP algorithm. 

In order to give a more quantitative comparison between our TEBD method and single site TDVP, we also studied the numerical convergence and runtime of the different simulations. To properly address the problem of numerical convergence, one has to carefully consider the two possible sources of error arising in our time evolution algorithm. From one side, both approaches possess a deterministic source of error coming from their respective truncation schemes, which is of order $\delta^2$. On the other hand, a second source of error is represented by the truncation of the bond dimension of the MPS. Indeed, at each time step of our TEBD algorithm we need to apply the MPO representing the time evolution operator to the MPS representing the state at time $t$. This generates a new MPS, corresponding to the state at time $t+\delta$, whose maximum bond dimension is the product of the bond dimensions of the MPO and the original MPS. Therefore, to avoid its exponential growth during the time evolution, the bond dimension needs to be truncated at each time step. This can be done using standard techniques, such as SVD or QR truncations or by performing a variational MPO-MPS product~\cite{Paeckel2019}. In this work, we employed the latter technique, as it produces better results at a smallest computational cost. In particular, at the beginning of the time evolution, we choose a maximum bond dimension $\chi$ for the MPS, which will be kept constant during the time evolution, similarly to what is done in a single site TDVP approach. Then, at each time step, a variational algorithm~\cite{Paeckel2019} is used to find the MPS which best fits the product of the time evolution operator MPO and the MPS at the previous time. In order to estimate the errors arising from the truncation of the MPS bond dimension, one can then repeat the simulation of the time evolution for different values of $\chi$, and study the convergence of the algorithm in terms of this parameter. Of course, while the procedure that we just described corresponds to the simplest approach, one could also think of varying $\chi$ along the time evolution. However, since the best implementation of such scheme might vary from case to case, we will limit ourselves to showing results for the case in which $\chi$ is kept constant during the evolution. In particular, in Fig.~\ref{fig:annni_err} we plot the differences between data collected for the bipartite entanglement entropy at bond dimensions $\chi$ and $\chi-10$ for different values of $\chi$, which represent an estimate of the error committed by truncating the bond dimension of the MPS. As one can see, the bond dimension truncation errors can be easily be made smaller than the trotterization ones. Moreover, we would like to highlight that not only the errors committed using our method are at least one order of magnitude smaller than those observed with TDVP for $J_2=0.15$ and $\chi=300$, but also the simulation times are an order of magnitude shorter (see Fig.~\ref{fig:annni_err}). This confirms that the TEBD approach proposed in this manuscript represents a valid alternative to other well established techniques. For the rate function of the Loschmidt echo we obtained, for different values of $J_2$, errors of the order of $10^{-6}$ - $10^{-8}$ for similar bond dimensions. This signals, with no huge surprise, that for larger systems (we switched from $N=60$ to $N=100$) we would need to increase the bond dimension to further reduce the truncation errors.

\subsection{2D Quantum Ising cylinder}
As a next step, we considered a two-dimensional model, namely the quantum Ising model on a cylinder. The Hamiltonian of the model is given by
\begin{align}
\label{eq:ising-ham}
   H &=  \sum_{\ell_x,\ell_y=1}^{L_x, L_y-1} (J_x\sigma^x_{\ell_x,\ell_y} \sigma^x_{\ell_x+1,\ell_y}+J_y\sigma^x_{\ell_x,\ell_y} \sigma^x_{\ell_x,\ell_y+1})\nonumber\\&+h\sum_{\ell_x,\ell_y=1}^{L_x, L_y}\sigma^z_{\ell_x,\ell_y},
\end{align}
where $J_x$ and $J_y$ parametrize the interactions along the two spatial directions, $h$ is a transverse field and periodic boundary conditions are applied only along the horizontal direction. 

\begin{figure*}[th!]
    \centering
    \includegraphics[width=\textwidth]{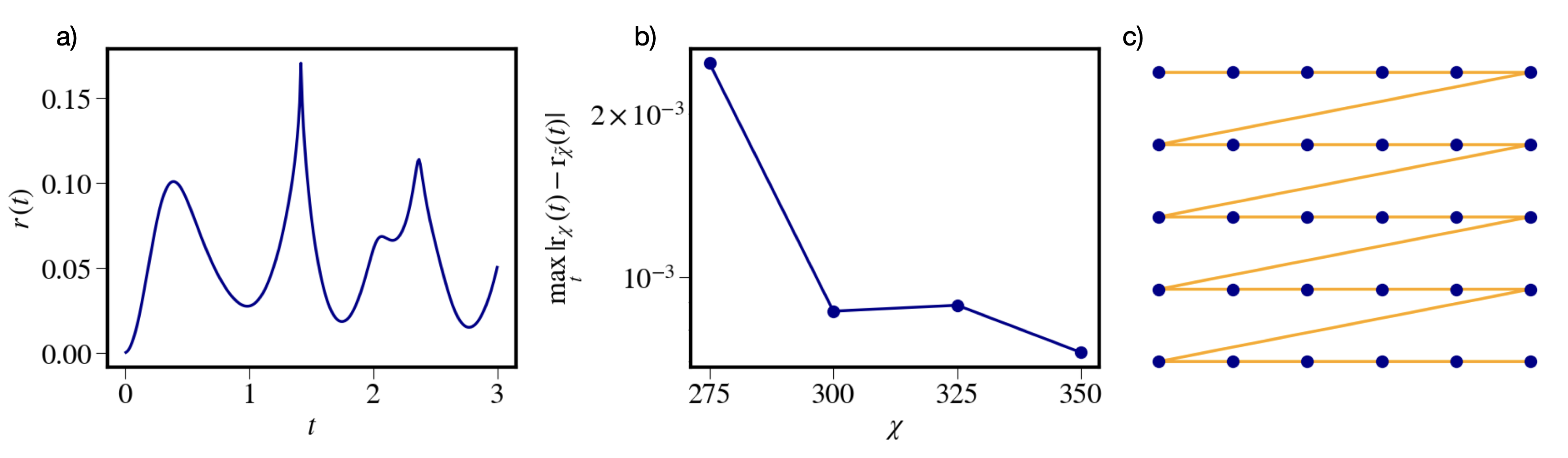}
    \caption{a) Evolution of the rate function of the Loschmidt echo after a global quantum quench in the external magnetic field from $h_0=0$ to $h_1=1.3$. We considered $J_x=J_y=1$, $L_x=6$, $L_y=50$ and a time step of $\delta=0.01$. b) The right panel, instead, shows the scaling of the truncation error with the bond dimension $\chi$. c) Zigzag mapping from two-dimensional to one-dimensional lattice.}
    \label{fig:ising2D}
\end{figure*}

To apply the MPS formalism to simulate the time evolution of this model, we first map the system to a one-dimensional chain with long-range interactions using the zig-zag mapping shown in Fig.~\ref{fig:ising2D}. Since, with this mapping, we will have sites interacting at a distance of six lattice spacings, at a first glance one might think that we will observe an drastic increase in the bond dimension of the trotterized time evolution operator. However, using our approach for the exponentiation of Pauli matrices, we can easily encode this operator as a short sequence of MPOs with small bond dimensions. Indeed, we can decompose the total time evolution operator as
\begin{equation}
    U^{TEBD2}=U_z U_x \bigg( \prod_{i=1}^{L_x}U_y^{(i)} \bigg)U_z,
\end{equation}
where $U_z$ has bond dimension $w=1$ and encodes the exponential of the single site operators, $U_x$ has bond dimension $w=4$ and includes all the interactions along the horizontal direction, and $U_y^{(i)}$ has bond dimension $w=2$ and encodes the interactions along the $i$-th column. Therefore, to time evolve our MPS state, we will need to perform (upon absorption of the trivial $U_z$ terms inside the other MPOs at the boundaries) $L_x+1$ MPO-MPS products, but each of the MPOs possesses a very small bond dimension.
    
Also in this case we studied the dynamical behavior of the rate function of the Loschmidt echo after a global quantum quench in the external field, and the results are reported in Fig.~\ref{fig:ising2D}. We considered a system of $N=300$ spins ($L_x=6$, $L_y=50$), and our results are in perfect agreement with those reported in~\cite{Hashizume2022}, with a truncation error $<10^{-3}$ for MPS bond dimension of $\chi=350$. It is reasonable to expect, however, that this error could be reduced at the cost of further increasing the bond dimension of the system, or by introducing a different 2D to 1D mapping, such as the Hilbert curve~\cite{Cataldi2021}.

\subsection{Other approximation schemes}
Finally, to further highlight the adaptability of our method, we will show that it can be easily applied to truncations of the unitary evolution operator which differ from the Trotter-Suzuki decomposition~\eqref{eq:Trotter}. To make a concrete example, let us conshider again the Hamiltonian in~\eqref{eq:model}
\begin{align}
    H_r(J,h)&=J\sum_{l=1}^{N-r}\sigma_l^x\sigma_{l+r}^x + h\sum_{l=1}^N\sigma_l^z \nonumber\\&= J X_{r,N} + hZ_{N}. 
\end{align}
Using the Zassenhaus formula~\cite{Magnus,Kleinert2004} for the exponential of the sum of two operators, we have that
\begin{align}
\label{eq:Zassenhaus}
    e^{-i\delta H_r}&=e^{-iJ\delta X_{r,N}}e^{-ih\delta Z_{N}}e^{\frac{J h \delta^2}{2}[X_{r,N},Z_N]}+o(\delta^3)\nonumber \\&=U^{Z2}+o(\delta^3).
\end{align}
\begin{figure}[t]
    \centering
    \includegraphics[width=.9\columnwidth]{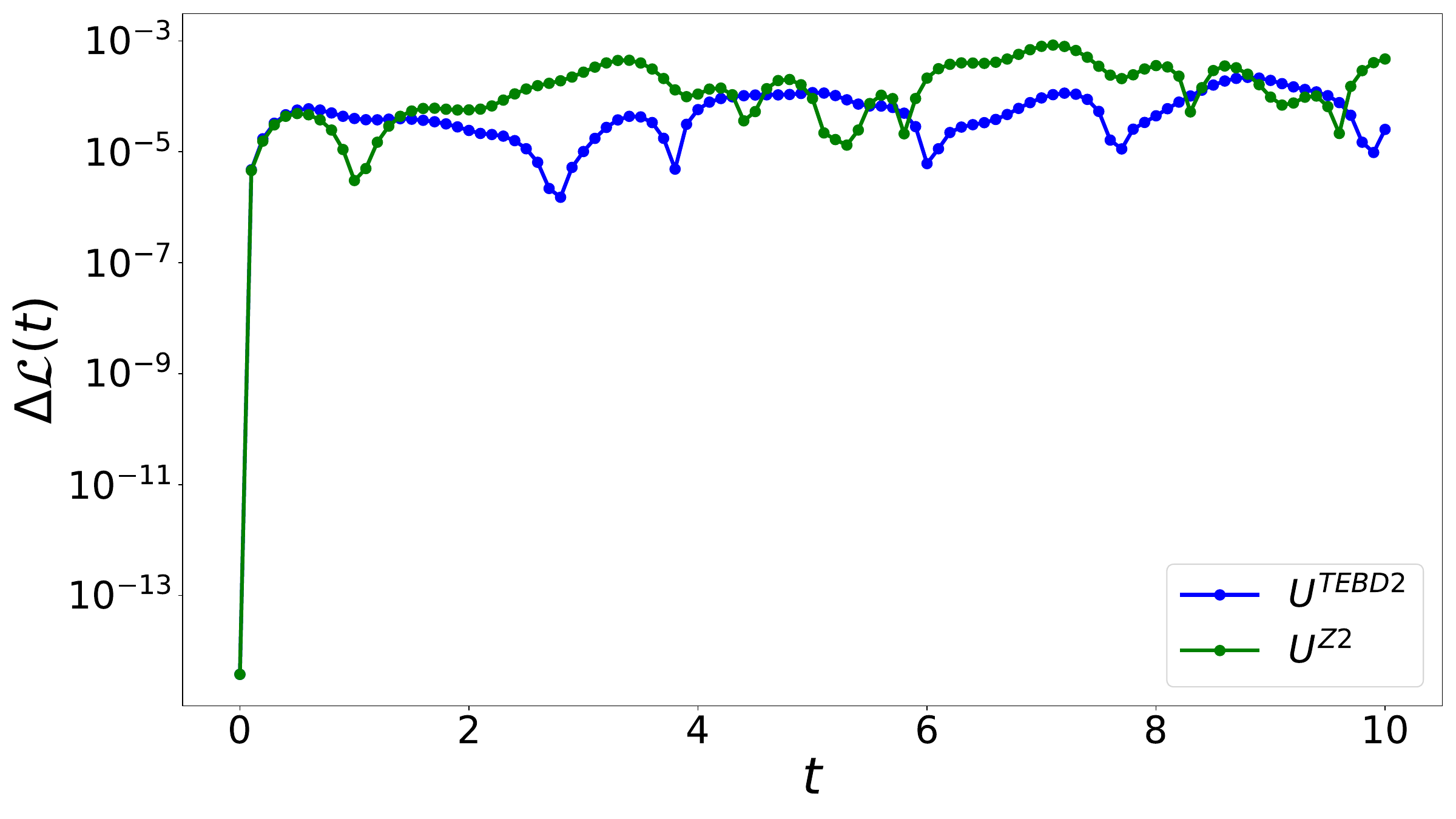}
    \caption{Error in the Loschmidt echo using different truncation schemes, measured with respect to exact diagonalization. Data are obtained for Hamiltonian~\eqref{eq:model} with $r=3$, $N=15$, $J=1$ and $h=0.3$, with a quench amplitude of $\Delta h=0.5$. The time step used for the time evolution is $\delta=0.01$. Blue dots represent error obtained with TEBD2 while the green line corresponds to exact that obtained with $U^{Z2}$. }
    \label{fig:Error}
\end{figure}
Therefore, the approximation of the time evolution operator with $U^{Z2}$ yields to a truncation error of the same order as $U^{TEBD2}$. The commutator appearing in~\eqref{eq:Zassenhaus} is given by the sum of two-qubit operators
\begin{equation}
    [X_{r,N},Z_N]=-2i\sum_{l=1}^{N-r}(\sigma_l^y\sigma_{l+r}^x+\sigma_l^x\sigma_{l+r}^y),
\end{equation}
whose exponential can be easily evaluated using our method after a first order trotterization, which only yields an error of order $\delta^4$, leaving unaltered the leading order given by~\eqref{eq:Zassenhaus}. In Fig.~\ref{fig:Error} we show the errors obtained when computing the Lodschmidt echo during the time evolution of the ground-state of~\eqref{eq:model}, for $r=3$, after a global quantum quench in the transverse magnetic field, using both $U^{Z2}$ and $U^{TEBD2}$. As expected, the errors of the two truncation schemes are of the same order of magnitude. 

To reduce the error one could truncate the expansion~\eqref{eq:Zassenhaus} to higher order, which will require the calculation of higher order nested commutators. These, however, will produce other Pauli strings, whose exponential is easily computed within our approach. The main drawback of going to higher orders in the expansion, as in TEBD4, would be that more MPO-MPS contractions and subsequent compressions are required. 
Nonetheless, we would like to stress that if one is able to think of any truncation scheme which will reduce the error per time step, and this scheme involves the exponentiation of Pauli matrices, our method ensure the construction of compact MPOs for such operators. 
\section{Conclusions}
\label{sec:conclusions}
In this work we have proposed an alternative technique for the construction of MPOs for the exponentials of non-local spin operators. This technique is based on the direct exponentiation of Pauli matrices, and finds its natural application in systems with long-range interactions, periodic boundary conditions and cluster interactions. The main advantage of this method is that the maximum bond dimension of the MPOs scales as $2^r$ if $r$ is the range of the interaction, which despite the exponential scaling provides better perfomances with respect to the standard approach. This renders the study of the dynamics of quantum many-body systems with TEBD more efficient, in terms of MPO-MPS contraction and MPOs size. Moreover, this technique is highly versatile, and provides a very natural way of exponentiating any spin interaction, producing MPOs containing only single site tensors without needing any additional manipulation. We tested the technique on some non-integrable models, measuring the time evolution of the Loschmidit echo and of the half-chain entanglement entropy. In all cases, we were able to reproduce known results about entanglement growth and dynamical phase transitions in these systems, taking into account systems of up to $N=300$ spins. Moreover, our analysis shows that our TEBD approach can reach better accuracies than other competitive methods such as single site TDVP. Finally, we gave an example of how our approach can be easily adapted to truncation schemes which are different from the Suzuki-Trotter decomposition, therefore it would be interesting to explore the possibility of applying it also to other approximation schemes for the unitary time evolution operator, such as the one recently proposed in \cite{Schilling2024}. Interestingly, the approach can also be applied to the simulation of the time evolution of realistic Rydberg atoms systems, and beyond time evolution problems it can find potential applications to the simulation thermal states and quantum circuits. Indeed, the same reasonings that we applied to trotterized time evolution operators can be extended to two qubit gates. We leave the exploration of these possibilities to future works.  
\section*{Acknowledgements}
I thank F. Franchini and S. M. Giampaolo for insightful discussion. A. G. C. acknowledges support from the MOQS ITN programme, a European Union’s Horizon 2020 research and innovation program under the Marie Sk\l{}odowska-Curie grant agreement number 955479.
The research leading to these results has received funding from the following organizations: project EuRyQa (Horizon 2020); Italian Ministry of University and Research (MUR) via: Quantum Frontiers (the Departments of Excellence 2023-2027); the World Class Research Infrastructure - Quantum Computing and Simulation Center (QCSC) of Padova University; Istituto Nazionale di Fisica Nucleare (INFN): iniziativa specifica IS-QUANTUM. 
\bibliography{biblio}
\end{document}